\documentclass[useAMS,usegraphicx,usenatbib]{mn2e}

\usepackage{graphicx}
\usepackage{txfonts}

\title[GC chemical evolution]{Chemical evolution of the Galactic Center}
   \author[V.~Grieco et al.]{V.~Grieco,$^{1,2,3}$\thanks{E-mail: 
       grieco@oats.inaf.it} F.~Matteucci,$^{1,2,3}$ N. Ryde,$^{4}$ M. Schultheis, $^{5}$  S. Uttenthaler$^{6}$\\
     $^1$Dipartimento di Fisica, Sezione di Astronomia, Universit\`a di 
     Trieste, Via G.B. Tiepolo 11, I-34143 Trieste, Italy \\
     $^2$INAF, Osservatorio Astronomico di Trieste, Via G.B. Tiepolo 11, 
     I-34143 Trieste, Italy\\
     $^3$INFN, Sezione di Trieste, Via A. Valerio 2,
     I-34127 Trieste, Italy\\
     $^{4}$Department of Astronomy and Theoretical Physics, Lund Observatory, Lund University, Box 43, 221 00, Lund, Sweden e-mail:ryde@astro.lu.se\\
     $^{5}$Observatoire de la C\^ote d'Azur, CNRS UMR 7293, BP4229, Laboratoire Lagrange, 06304 Nice Cedex 4, France  e-mail: mathias.schultheis@oca.eu\\
     $^{6}$University of Vienna, Department of Astrophysics, T\"urkenschanzstra\ss e17, 1180 Vienna, Austria}

\begin{document}

     \date{Accepted . Received ; in original form \today} 

     \pagerange{\pageref{firstpage}--\pageref{lastpage}} \pubyear{2013}

     \maketitle

     \label{firstpage}

\begin{abstract} In recent years, the Galactic Center (GC) region  (200 pc in radius) has been studied in detail 
with spectroscopic stellar data as well as an estimate of the ongoing star formation rate.
The aims of this paper are to study the chemical evolution of the GC region by means of a detailed chemical 
evolution model and to compare the results with high resolution spectroscopic data in order to impose constraints 
on the GC formation history.The chemical evolution model assumes that the GC region formed by fast infall of 
gas and then follows the evolution of $\alpha$-elements and Fe. 
We test different initial mass functions (IMFs), efficiencies of star formation and gas infall timescales. 
To reproduce the currently observed star formation rate, we assume a late episode of star formation triggered 
by gas infall/accretion. We find that, in order to reproduce the [$\alpha$/Fe] ratios as well as the metallicity 
distribution function observed in GC stars, the GC region should have experienced a main early strong burst of 
star formation, with a star formation efficiency as high as $\sim 25 \, Gyr^{-1}$, occurring on a timescale in the 
range $\sim$ 0.1-0.7 $Gyr$, in agreement with previous models of the entire bulge. 
 Although the small amount of data prevents us from drawing firm conclusions, we suggest that the best 
IMF should contain more massive stars than expected in the solar vicinity, and the last episode 
of star formation, which lasted several hundred million years, should have been triggered by a modest episode 
of gas infall/accretion, with a star formation efficiency similar to that of the previous main star 
formation episode. This last episode of star formation produces negligible effects on the abundance patterns 
and can be due  to accretion of gas induced by the bar. Our results exclude an important infall event as a 
trigger for the last starburst.
\end{abstract}


\section{Introduction}
In the last years there has been an increasing interest in studying the Galactic bulge resulting from the accumulation  of more spectroscopic data.
In particular, from these data a rather complex picture for the bulge formation arose: the existence of at least two main stellar populations 
in the bulge was  suggested by several studies (Babusiaux et al. 2011; Gonzalez et al. 2011; Hill et al. 2011; Bensby et al. 2011, 
Robin et al. 2012, Uttenthaler et al. 2012; Bensby et al. 2013). The two populations arising from the spectroscopic data are characterized
by a metal-poor one, with characteristics typical of stars belonging to a spheroid, and another one more metal-rich and with 
characteristics of bar kinematics. Also the [$\alpha$/Fe] ratios of these two populations seem different in the sense that the metal 
rich population shows lower  [$\alpha$/Fe] ratios (Hill et al. 2011).
Grieco et al. (2012, hereafter GMPC12) tried to model these two populations and concluded that the metal-poor population  formed on a 
very short timescale by means of an intense burst of star formation, whose efficiency was a factor of 20 higher than in the Galactic 
thin disk. The more metal-rich population instead should have formed on a longer timescale and with lower star formation efficiency 
and it was created by the bar evolution. While in the GMPC12 model the bulge was treated as a unique single zone with radius of 2kpc, 
here we present a model for the central region of the bulge, the Galactic Center (GC). In particular we will focus on a region of 200 
pc radius. The GC region has an extremely rich environment: intense star formation with formation of a bulk of massive stars 
and three of the most massive young star clusters in the Galaxy. 
In the last years many assumptions have been made about the nature of the inner population and on the possible causes of the star 
formation activity in the last few hundreds Myr. Genzel et al. (2006)  pointed out that most of the stars in the GC seem to have formed  
$9 \pm 2$ $Gyr$ ago, probably at the same time as the Galactic bulge. Then the star formation rate (SFR) dropped to a minimum a few $Gyr$
ago rising again in the past few million years. The star formation history was derived in the GC from  the H-R diagram of the red giants 
(Pfhul et al. 2011), while the present time SFR was derived from the infrared luminosity of the young massive stars forming in the large 
reservoir of molecular gas present in the GC (Yusef-Zadeh et al. 2009) or by counting these young massive stars (Immer et al. 2012).
One important issue relevant to the nature of the stellar populations in the GC is the possible presence of a second bar.
However, the presence of a double bar in our Galaxy, similar to what can be observed in external galaxies (Laine et al. 2002; Erwin 2004)
is still under debate. Alard (2001), Nishiyama et al. (2005), and Gonzalez et al. (2011) claim that there is an inner structure distinct
from the large-scale Galactic bar, with a different orientation angle which could be associated with a secondary, inner bar. On
the other hand, Gerhard \& Martinez-Valpuesta (2012) can explain this inner structure by dynamical instabilities from the disk
without requiring a nuclear bar. 
Very recently, Ryde \& Schultheis (2015) have measured, from high resolution infrared spectroscopy, abundances of Mg, Si, Ca and Fe 
in 9  M-type giants in the GC region.  They found a metal-rich population  at [Fe/H]=+0.11$\pm$0.15 dex with rather low [$\alpha$/Fe] ratios 
with the possible exception of Ca. Their metallicity distribution function (MDF) is peaked at a [Fe/H] slightly higher than solar and 
agrees with previous data of Cunha et al. (2007). Cunha et al. (2007) had also measured high resolution infrared spectra for a sample 
of cool stars within 30 pc from the GC; their results showed that [O/Fe] and [Ca/Fe] are enhanced by 0.2 and 0.3 dex, respectively.
They also found that the total range in [Fe/H] in the GC is narrower than that found for the older bulge population.

The aim of this paper is to compare our results with the data of Ryde \& Schultheis (2015) in order to impose constraints on the 
formation and evolution of the GC. The analysis of the abundance ratios, such as [$\alpha$/Fe] ratios,
versus [Fe/H] can allow us to understand the timescale of the process of formation of the GC region. In fact, [$\alpha$/Fe] ratios 
can be used as cosmic clocks because of the time-delay model. The time-delay model is the common interpretation of abundance patterns 
in terms of different chemical elements produced from different stars and on different timescales. In this context, starting from the 
fact that $\alpha$-elements are produced on short timescales by core-collapse SNe,  whereas Fe is produced partly in core-collapse SNe 
but mainly in Type Ia SNe exploding on longer timescales,  we can interpret the observed abundances. High [$\alpha$/Fe] ratios at low [Fe/H] 
indicate that the star formation period was short enough to prevent the Type Ia SNe from  polluting heavily the gas with Fe. On the other hand, 
we always expect that the [$\alpha$/Fe] ratios are lower at high [Fe/H] because of Type Ia SN contribution. However, due to the 
combinationation of the star formation rate (SFR) and the time-delay in element 
production, objects where the SFR has been very intense show high [$\alpha$/Fe] ratios up to even solar [Fe/H] values. This is 
because $\alpha$-elements and part of Fe are produced  fast by core-collapse SNe, which are able to enrich substantially the ISM 
in Fe before the bulk of this element is restored by SNe Ia (see Matteucci, 2012).  Therefore, the comparison between observed and 
predicted abundances can tell us about the history of the chemical evolution of the GC and in particular on the origin of its stellar 
populations.
To perform  such a comparison, we adopt a model very similar to that presented in GMPC12, in particular the model describing the 
classical bulge population which formed on a short timescale as a result of an intense burst of star formation (the population called metal-poor), 
applied to a central region of only 200 pc of radius. In order to reproduce the star formation observed at the present time, we also 
overimpose on the history of star formation, deriving from our model, a recent burst producing the same quantity of stars that have been observed.

The paper is organized as follows: in section 2 we describe the chemical evolution model, in section 3 we describe the stellar spectroscopic 
data as well as the SFR estimates. In section 4 we present our results and the comparison with data. In section 5 we draw some conclusions.

\section{The chemical evolution model}
The chemical evolution model we adopt here is the model of GMPC12 for the metal-poor (MP) stellar population in the Galactic bulge. 
In other words, it is the model assuming a strong initial burst of star formation triggered by the heavy gas infall  
coming from the initial collapse that gave rise to the halo and the bulge. This model has been adapted here to the small 
central region around the Galactic Center (GC).
The assumed gas accretion law is:
\begin{equation}
\dot{\sigma}_{gas, i}(t)=A(r) X_{i, inf}e^{-t/T_{inf}}
\end{equation}
where $i$ represents a generic chemical element, $T_{inf}$ is an appropriate infall timescale fixed by reproducing the 
observed stellar metallicity distribution function,  and $A(r)$ is a parameter constrained by the requirement of 
reproducing the current total surface mass density in the central region of the Galactic bulge; in particular, we assume a present 
time surface mass density at $r=200$ pc,
$\sigma_{GC}(200,t_G) \sim 3.9 \cdot 10^{5} M_{\odot}pc^{-2}$, with $t_G \sim$14 $Gyr$ being the lifetime of the Galaxy, and $r$ the 
Galactocentric distance. We have obtained this value by assuming a mass distribution in the bulge which follows a Sersic (1968)  profile.
The quantity $X_{i, inf}$ is the abundance of the element $i$ in the infalling gas.
The parameter $T_{inf}$ will be allowed to vary in order to find the best agreement with the present time observables (abundances, SFR and stellar mass).
The star formation rate (SFR) is a Schmidt- Kennicutt law:
\begin{equation}
\dot{\sigma}_{gas}= \nu \sigma_{gas}^{k}
\end{equation}
with $k=1$ as in previous papers (Ballero et al. 2007; Cescutti \& Matteucci, 2011; GMPC12). 
The quantity $\nu$ is the efficiency of star formation, namely the SFR per unit mass of gas, and for the bulge is assumed to be 
 high relative to that normally assumed in the solar vicinity ( $\nu=1 \,Gyr^{-1}$, see  for example Chiappini et al. 1997), thus simulating a star burst. 
The efficiency of star formation is in units of $time^{-1}$ and represents the inverse of the time scale in which all the gas is consumed. 
This approach was first suggested by Matteucci \& Brocato (1990) and then proven to be
required for reproducing the bulge properties in all the following papers dealing with the chemical evolution of the bulge. 
Here we assume several values for this parameter varying from $\nu$ from 20 to 200 $Gyr^{-1}$.

We explore different initial mass functions (IMFs), in particular:  
\begin{itemize}
 \item[i)] 
 the one suggested by Ballero et al. (2007):
\begin{equation}
\phi_{B}(M)=A_B M^{-(1+x)}
\end{equation}
with $x=0.95$ for $M > 1 M_{\odot}$ and $x=0.33$ for $M < 1
M_{\odot}$ in the mass range $0.1-100M_{\odot}$ and $A_B=0.078$.    
 \item[ii)] The normal Salpeter (1955) IMF (x=1.35 in eq. (3)) in the mass range  $0.1-100M_{\odot}$, with $A_S=0.17$.
 \item[iii)] The three-slope Kroupa et al. (1993) IMF:  
\begin{equation}
\phi_{K}(M) =  
\left\{
\begin{array}{rl}
A_{K}\, M^{-0.3} & \mbox{if} \,M < 0.5 M_{\odot} \\
B_{K}\, M^{-1.2} & \mbox{if} \, 0.5 < M/M_{\odot} < 1 \\  
C_{K}\, M^{-1.7} & \mbox{if} \,M > 1 M_{\odot}  
\end{array}
\right.
 \end{equation}
with $A_{K}\approx 0.19, \, B_{K}=C_{K}\approx 0.31 $.
 \item[vi)] The Chabrier (2003) IMF:
\begin{equation}
\phi_{C}(M) =  
\left\{
\begin{array}{ll}
A_{C} \, exp^{-(logM-logM_c)^{2}/2\sigma^{2}} & \mbox{if} \,M < 1 M_{\odot} \\
B_{C} \, M^{-1.3} & \mbox{if} \, M > 1 M_{\odot}
\end{array}
\right.
 \end{equation}
with $A_{C}\approx 0.85, \, B_{C}\approx 0.24, \, \sigma=0.69 \,\, \mbox{and} \,\, M_c=0.079 $. 
\end{itemize}

We note that this IMF is very similar to that of Kroupa (2001).

 \subsection{Nucleosynthesis and stellar evolution prescriptions}
 
The nucleosynthesis prescriptions adopted here are different from those in Grieco et al. (2012) and are the 
same as in Romano et al. (2010): they take into account the most recent calculations and have been tested 
on the chemical evolution of the solar neighbourhood and found to be the best ones.

In particular, the prescriptions for different stellar mass ranges are:
\begin{itemize}
\item 
 the yields of Karakas (2010) for low and intermediate mass stars  ($0.8\le M/M_{\odot} \le8$, LIMS),
 who has recomputed Karakas \& Lattanzio's (2007) models with 
 updated nuclear reaction rates and enlarged the grid of masses and metallicities.
\item The yields for massive stars ($M>8M_{\odot}$) are taken from Kobayashi et al. (2006), which included
metallicity-dependent mass loss, plus the yields of the Geneva group for C, N, O
(Meynet \& Maeder 2002a; Hirschi et al. 2005; Hirschi 2007; Ekstrom et al. 2008), limited to the presupernova stage, but
computed with both mass loss and rotation.
\item  The yields from Type Ia SNe are assumed to originate from exploding white dwarfs in binary systems and are taken from
Iwamoto et al. (1999). The SN Ia rate is computed as in the previous papers (Ballero et al. 2007; Cescutti \& Matteucci, 2011; Grieco et al. 2012) 
by assuming the single-degenarate model, as described in Greggio \& Renzini (1983) and Matteucci \& Recchi (2001). 
This particular formulation,   namely the single-degenerate scenario, gives results very similar to the double-degenerate scenario for the progenitors of SNe Ia, 
as shown by Matteucci et al. (2006; 2009).
\end{itemize}

\section{Observations} 
\subsection{Abundances}
The nine M-type giants in the Galactic-center region discussed here were analysed in Ryde \& Schultheis (2015) using high spectral 
resolution in the K band, centred at 2.1 $\mu$m. They were selected using the Nishiyama et al. (2009)
dataset and the 3D interstellar extinction maps of Schultheis et al. (2014) to ensure that these giants are situated in the Galactic center region.
The low-resolution K band spectra were used to determine the effective temperatures ($T_\mathrm{eff}$) with a precision of approximately $100\,\mathrm{K}$. 
The surface gravities ($\log g$),  were determined by adopting a mean distance of 8 kpc and using dereddened H and Ks magnitudes from the 
VVV survey of Minniti et al. (2010), resulting in a precision of less than 0.3 dex.  

The high-resolution spectra, centered at $2.1\,\mu$m, were used to determine the metallicity ([Fe/H]) from several Fe lines and the Ca, Si, 
and Mg abundances from several atomic lines. To derive the abundances, spectral synthesis was used with  the software  {\it Spectroscopy Made Easy (SME)}, 
Valenti \& Piskunov (1996,2012), based on MARCS spherical-symmetric, Local Thermodynamic Equilibrium (LTE) model atmospheres (Gustafsson et al. 2008).  
The abundance from an observed spectral line was found by minimising the $\chi^2$, calculated from the comparison between synthetic spectra and 
the observations. The best fit was found, with the abundance as a free parameter, by minimising the $\chi^2$. The microturbulence of 
$\xi_\mathrm{micro}=2.0\pm0.5\,\mathrm{km\,s^{-1}}$ was used in the analysis. 

The uncertainties, due to the uncertainties in the fundamental parameters ($T_\mathrm{eff}$, $\log g$,  [Fe/H], and  $\xi_\mathrm{micro}$), 
in the derived metallicities and abundance ratios are of the order of 0.10-0.15 dex. There might also be unknown systematic uncertainties, 
such as non-LTE effects, but since the standard red-giant star, Arcturus, is among the analysed stars and gives reasonable results, these 
uncertainties cannot be large. Unknown blends might also exist, especially for metal-rich stars, and will tend to overestimate the metallicities. 
However, the line blending is less severe in the K band compared to the optical region. Furthermore, since several Fe lines were used, this effect 
would rather lead to an increased dispersion in the abundances from the different lines. The molecules in the wavelength regions  are predominantly  
due to CN, and are properly synthesised.  

The metallicity distribution of Ryde \& Schultheis (2015) agrees well with that of Cunha et al. (2007). These authors also measured Ca abundances, 
which also agree well with those measured by Ryde \& Schultheis (2015).

\subsection{The Central Molecular Zone}

The understanding of the physical processes occuring in the nuclear disk of our Galaxy is crucial for the insight  into the formation and evolution of our own Milky Way. 
The Central Molecular  Zone (CMZ) is the innermost $\sim$ 200 arcmin region of the Milky Way.
It covers about $\rm -0.7\degr   < l < 1.8\degr$  in longitude and $\rm -0.3\deg < b < 0.2\degr$ in latitude which is within about 300\,pc of the  
Galactic Center.  It is  a  giant molecular cloud complex with an asymmetric distribution of molecular clouds (see e.g. Morris \& Serabyn 1996, 
Martin et al. 2004 and Oka et al. 2005).
Genzel et al. (2006) pointed out that most of the stars in the GC seem to have formed $9\pm 2$  $Gyr$ ago, probably at the same time as the Galactic bulge. 
Then the SFR dropped to a minimum a few $Gyr$ ago and increased again in the past few million years.
Recent Herschel observations (Molinari et al. 2011) show that the CMZ is a 100 pc elliptical and twisted ring of cold dust, with 3 
$\times$ 10$^7$ M$_{\odot}$ of gas. This prodigious reservoir of molecular gas is an active region of star formation, with evidence of starburst activity 
in the last 100.000 years (Yusef-Zadeh et al. 2009). The  gas pressure and temperature are higher in the CMZ than in the average 
disk, conditions that favour a larger Jeans mass for star formation and an initial mass function biased toward more massive stars. Furthermore, the presence 
of strong magnetic fields, tidal shear, and turbulence demonstrate that the conditions for star formation in the CMZ are significantly different from those 
in the Galactic disk (see Serabyn \& Morris 1996, Fatuzzo \& Melia 2009).
From the observational point of view, the present star formation rate can be determined by counting massive young stellar objects (Immer et al. 2012).

 In order to estimate bolometric luminosities from their flux densities at 15\, $\mu$m,  one has to correct for interstellar dust extinction (Schultheis et al. 1999). 
The total luminosity of the young stellar objects has then been converted to a total mass assuming a Salpeter IMF for masses in the range of 6 to $\rm 58\,M_{\odot}$, while 
the low-mass objects  ($\rm M < 6.3\,M_{\odot}$) have been converted using the Kroupa (2001) IMF.

In order to calculate the SFR, Immer et al. (2012) 
 assumed a typical life-time of 1\,Myr for the young stellar objects which leads to an average star formation rate of $\rm 0.08\,M_{\odot}\,yr^{-1}$. 
However, one has to take into account that the derived SFR are based on simple assumptions which could lead to significant errors.
Previous studies (Crocker 2011 and Yusef-Zadeh et al. 2009) published a range of values of $\rm 0.08-0.15\,M_{\odot}\,yr^{-1}$. 
This is consistent with other studies using different tracers of gas and star formation acitvity leading to a present SFR of  $\rm 0.04-0.15\,M_{\odot}\,yr^{-1}$ 
(Morris \& Serabyn 1996, Yusef-Zadeh et al. 2009, Molinari et al. 2011 and Immer et al. 2012) with a total gas mass in the order of  $\rm 3-7 \times 10^{7} M_{\odot}$.  
 Given the total amount of gas in the CMZ, the SFR is by a factor of $\rm >10$ lower than expected. This suppression of star 
formation could be due to the high turbulent pressure (see e.g. Kruijssen et al. 2014).
Pfhul et al. (2011) presented spatially resolved imaging and integral field spectroscopy for 450 cool giants within 1 pc from SgrA*. 
They derived the star formation rate of the nuclear cluster in the Galactic Center from the H-R diagram of the red giant population.
They found that the SFR dropped from an initial maximum roughly 10 $Gyr$ ago to a minimum  at 1-2 $Gyr$ ago and that increased again in the 
last few hundred Myrs. Their results imply that $\sim$80\% of the total stellar mass in the GC formed more than  5 $Gyr$ ago.

 \section{Results}
 \subsection{Abundances}
Two of the most important observational constraints to understand the most 
probable scenario for the bulge formation are the stellar metallicity distribution function (MDF) of stars and the [$\alpha$/Fe] ratios. 
Originally, Matteucci \& Brocato (1990) suggested that in order to fit the MDF of the
Bulge one should assume that it formed very quickly. The consequence of the fast formation is a particular [$\alpha$/Fe] trend with 
super-solar values being present for a large interval of [Fe/H]. 
In other words, while in the solar vicinity the ``knee'' of the [$\alpha$/Fe] ratios occurs roughly  at [Fe/H]$\sim$-1.0, in the bulge 
the knee appears at [Fe/H]$\sim$ 0. This is due to the fact that in a regime of intense star formation, 
core-collapse SNe from massive stars enrich the interstellar medium (ISM) very fast in Fe (although they are not the main producers 
of this element), so when SNe Ia which produce the bulk of Fe occur, the abundance of Fe in the ISM is already almost solar 
(see Matteucci 2001, 2012).

\begin{figure*}
	  \centering   
    \includegraphics[scale=0.5]{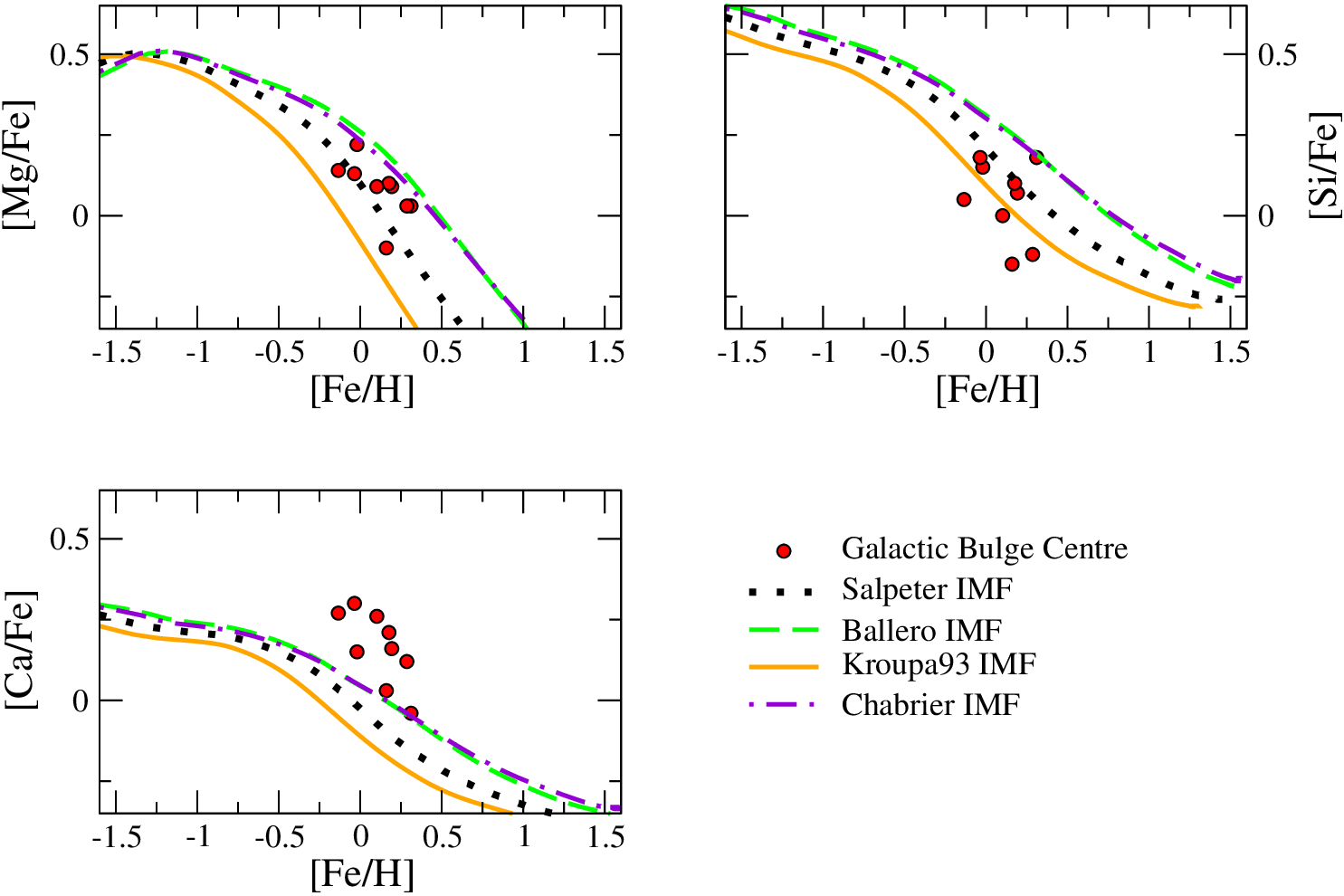} 
    \caption{Predicted [Mg/Fe], [Si/Fe], [Ca/Fe] vs. [Fe/H] for models with different IMFs, as indicated in the Figure. 
    The only parameter which varies from model to model is the IMF since each model has $\nu$=25 $Gyr^{-1}$ and $T_{inf}=0.7\, Gyr$.
The red filled circles are the abundances of Galactic Center stars from Ryde and Schultheis (2015).
    }
\label{MgSiCa_IMF}
\end{figure*}

We compare here the GC data with the chemical evolution model (hereafter the reference model) for a classical bulge of GMPC12. 
We computed several numerical models by varying the most important parameters: the IMF, the efficiency of star formation and the timescale of the infalling gas. 
The reference model parameters are: $\nu$=25 $Gyr^{-1}$, $T_{inf}$=0.1 $Gyr$ and the Salpeter IMF in agreement with the GMPC12G metal poor population. 
The data trend of Figure \ref{MgSiCa_IMF} shows the effect of an intense burst of star formation coupled with the delay in the Fe production from SNe Ia:
since star formation is very intense, the bulge very soon reaches a solar metallicity because of 
the core collapse SN explosions, which produce part of the Fe. After a time delay necessary to the evolution of the binary system progenitors of SNe Ia,
these SNe, which produce the bulk of Fe, start to explode changing the [$\alpha$/Fe] slope at a larger [Fe/H] than in the solar vicinity.
The models in Figure \ref{MgSiCa_IMF} show also the effect of varying the IMF: the Ballero IMF and the Chabrier IMF give very similar 
results since they are favouring massive stars  compared to the Kroupa et al. (1993)  and Salpeter (1955) IMFs.
 In particular, the Kroupa et al. (1993) IMF which has been derived for the solar vicinity does not reproduce either the [Mg/Fe] or the 
[Ca/Fe] ratios, although it could reproduce the [Si/Fe] abundances. On the other hand, the Salpeter (1955) IMF can reproduce the [Mg/Fe] and [Si/Fe] ratios. 
We note that none of the adopted IMFs can reproduce the high values of the [Ca/Fe] ratios. This could be due to the uncertainties still existing in 
the stellar yields of specific elements, but also to observational errors.
The abundance data are, in fact, more sensitive to the yield uncertainties rather than to the IMF, whereas the stellar MDF is highly sensitive to the IMF.

\begin{figure*}
	  \centering   
    \includegraphics[scale=0.5]{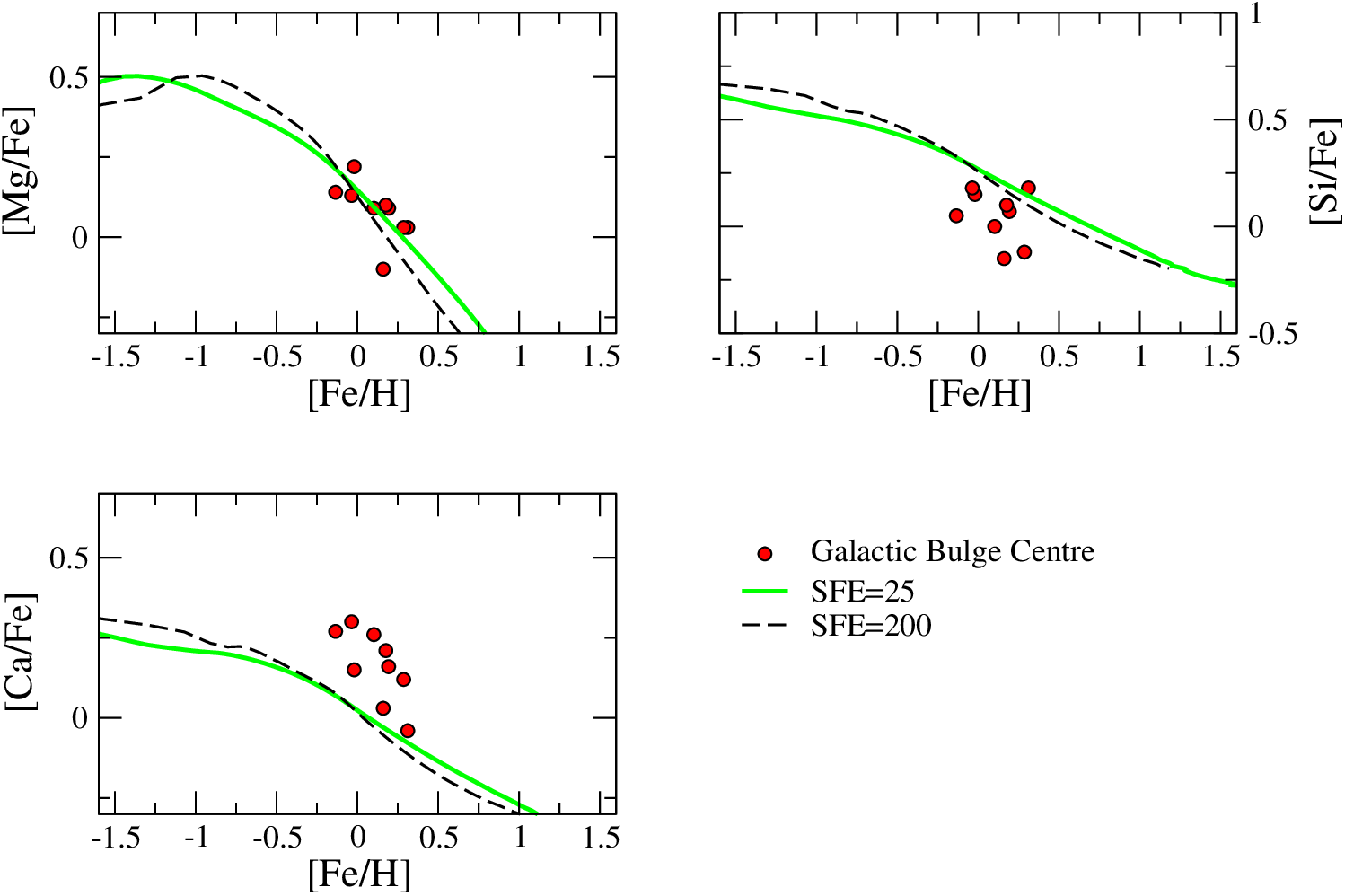} 
    \caption{Predicted [Mg/Fe], [Si/Fe], [Ca/Fe] vs. [Fe/H] by our fiducial model ($T_{inf}$=0.1 $Gyr$, Salpeter IMF) but with different star formation efficiency:
   $\nu=25\, Gyr^{-1}$ (green solid line) and $\nu=200\, Gyr^{-1}$ (black dashed line). 
  The red filled circles are the abundances of Galactic Center stars
    measurements from Ryde and Schultheis (2015), as in Figure 1.
    }
\label{MgSiCa_SFE}

\end{figure*}

In Figure \ref{MgSiCa_SFE} we can see the effects of a different star formation efficiency on the 
[Mg/Fe], [Si/Fe], [Ca/Fe] vs. [Fe/H] trends: 
the green solid line is the reference model while the black dashed line is a model with an intense initial 
starburst with efficiency $\nu$=200 $Gyr^{-1}$. 
This very high efficiency is probably unrealistic since it predicts that the gas is consumed in 5 Myr. 
In any case, the differences among different models 
with $\nu=25$ and 200 $Gyr^{-1}$ are negligible. On the other hand, 
a $\nu=1 Gyr^{-1}$ as in the solar vicinity 
(see Chiappini et al. 1997) would produce different results, in particular, 
all the curves would be translated at lower values of [$\alpha$/Fe] ratios.

\begin{figure*}
	  \centering   
    \includegraphics[scale=0.5]{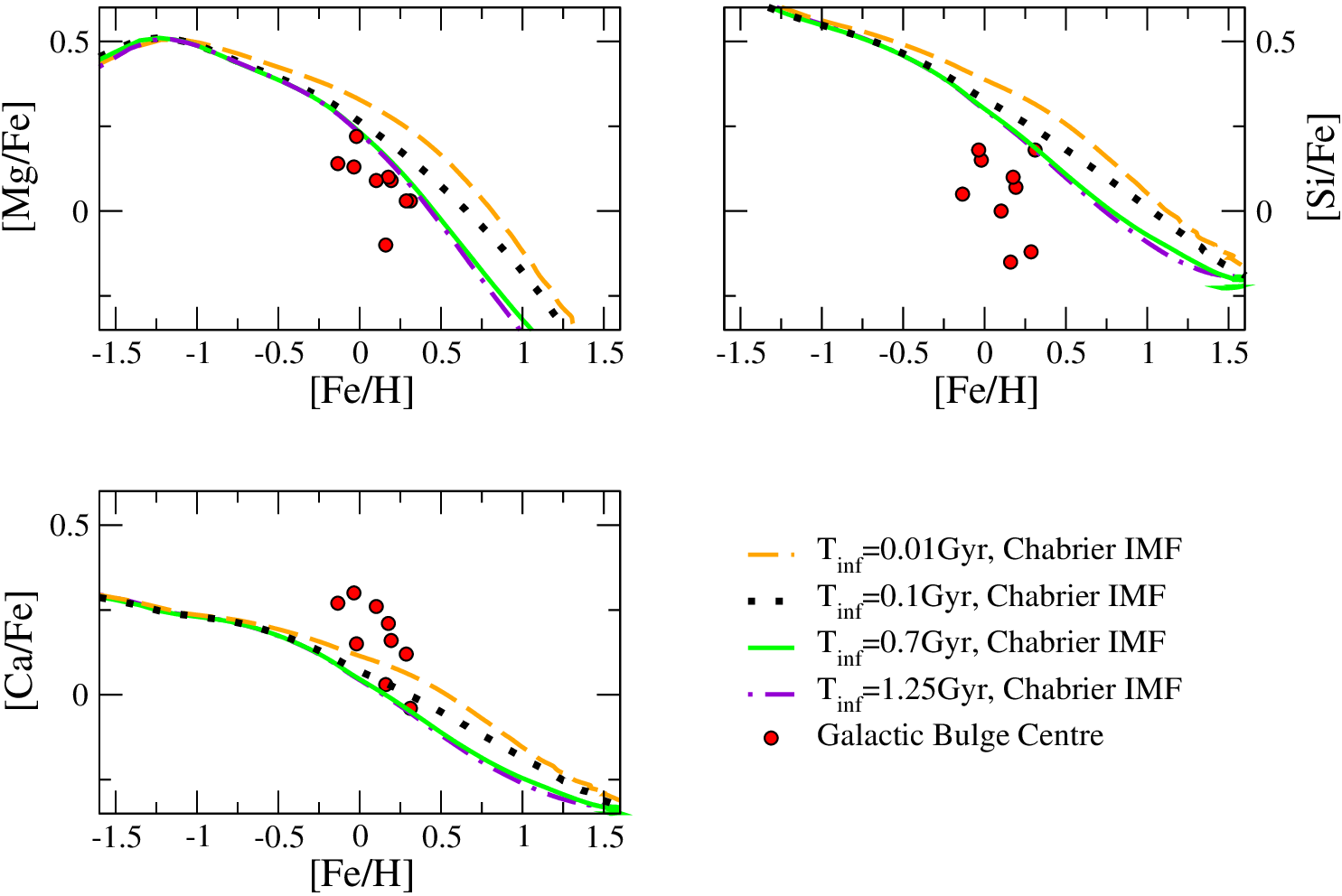} 
    \caption{Predicted [Mg/Fe], [Si/Fe], [Ca/Fe] vs. [Fe/H];  The only parameter which varies from model to model is the infall timescale
    in the range $T_{inf}$: 0.01-1.25  $Gyr$; we use the Chabrier IMF and the SFE of the fiducial model: $\nu$=25 $Gyr^{-1}$. 
    The red filled circles are the abundances of Galactic Center stars from Ryde and Schultheis (2015), as in Figure 1.
    }
\label{MgSiCa_Tinf}

\end{figure*}

In Figure \ref{MgSiCa_Tinf} we show the effect of varying the infall timescale.  
The differences among different models are small and  from the Figure we can conclude 
that a timescale between 0.1 and 1.25 $Gyr$ can be acceptable, whereas a timescale of 0.01 $Gyr$ seems too short.

\begin{figure*}
	  \centering   
    \includegraphics[scale=0.4]{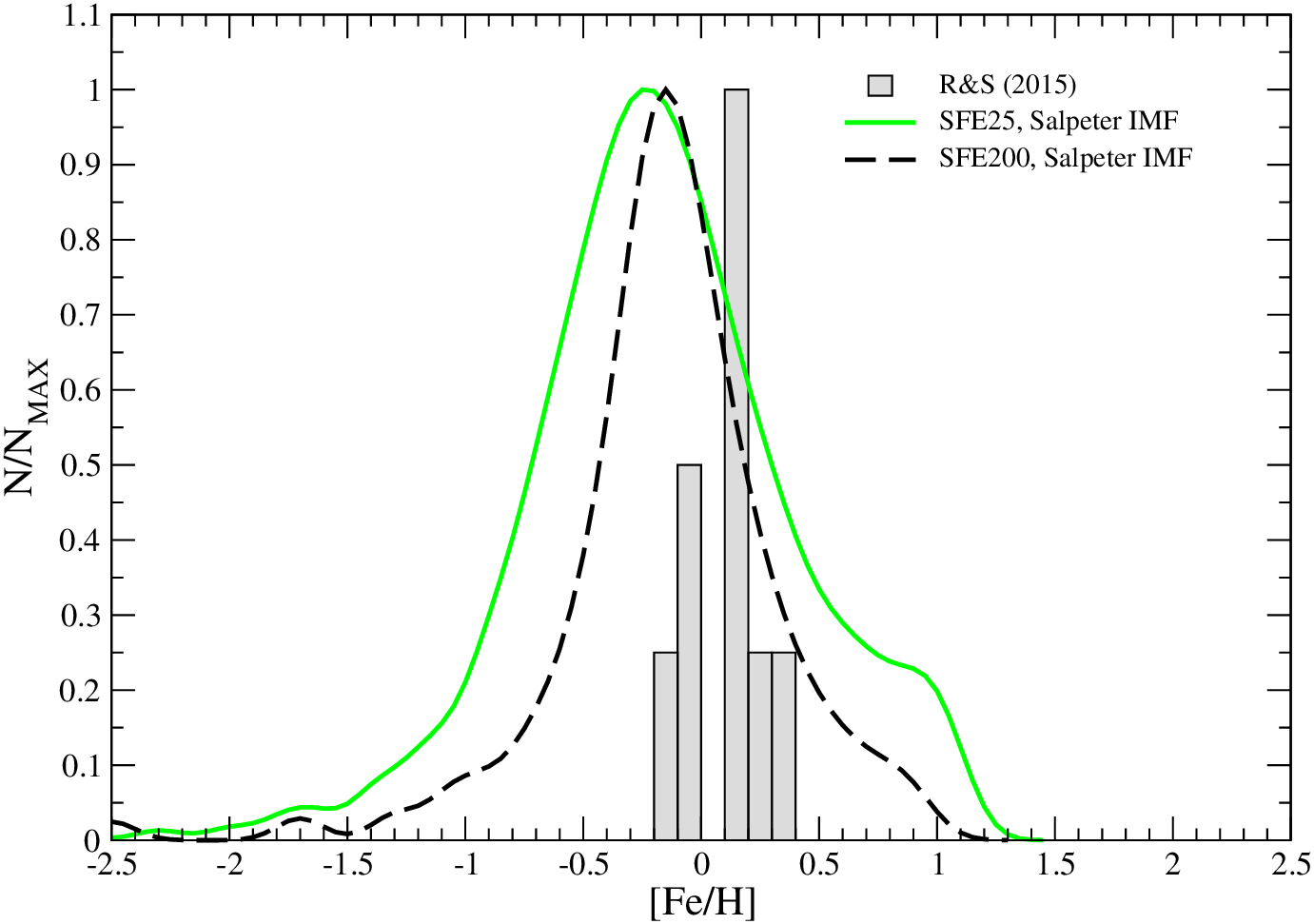} 
    \caption{Predicted MDF for the fiducial model (green solid line) and 
a model 
with SFE=$\nu$=200 $Gyr^{-1}$ (black dashed line) as a function of [Fe/H]. 
    The data are from Ryde and Schultheis (2015).  The predicted MDFs have been convolved with a Gaussian with an error of 0.1 dex.
    }
\label{MDF_SFE}
\end{figure*}

In Figure \ref{MDF_SFE} we show the predicted stellar MDF for the fiducial model and the two different 
values of $\nu$. Here, we can see that the effect of increasing $\nu$ results in a narrowing of the predicted MDF.  
It is worth noting that the number of observed stars in the GC is very small and therefore we do not aim at reproducing exactly 
the observed MDF but only at reproducing the position of the peak. 

\begin{figure*}
	  \centering   
    \includegraphics[scale=0.4]{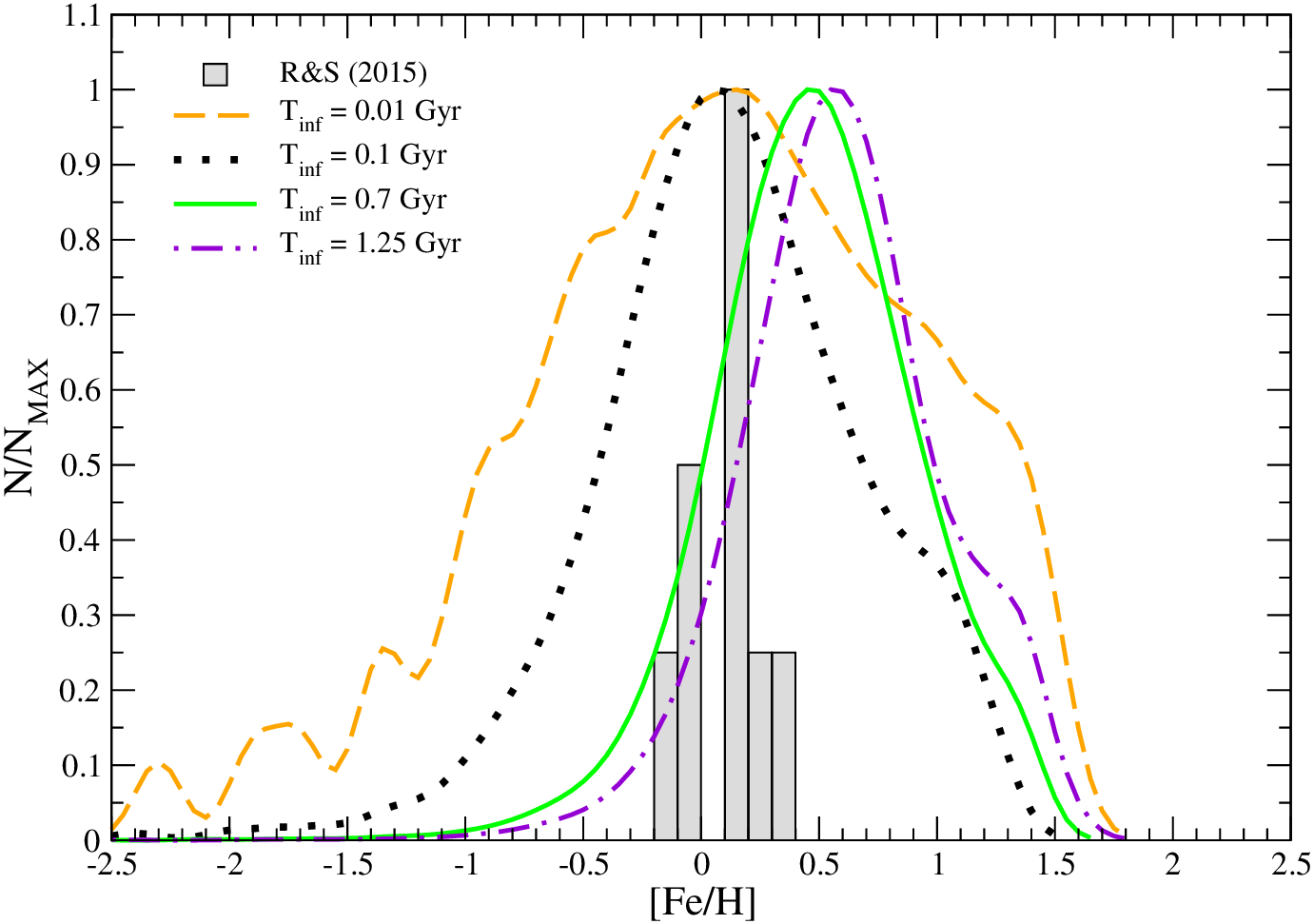} 
    \caption{Predicted MDF for different models with star formation efficiency SFE=$\nu=25\, Gyr^{-1}$ and the Chabrier IMF but 
    with different infall timescales. The lines have the same color code as in Figure \ref{MgSiCa_Tinf}.
    The data are from Ryde \& Schultheis (2015).  The predicted MDFs have been convolved with a Gaussian with an error of 0.1 dex.
    }
\label{MDF_Tinf}
\end{figure*}

\begin{figure*}
	  \centering   
    \includegraphics[scale=0.4]{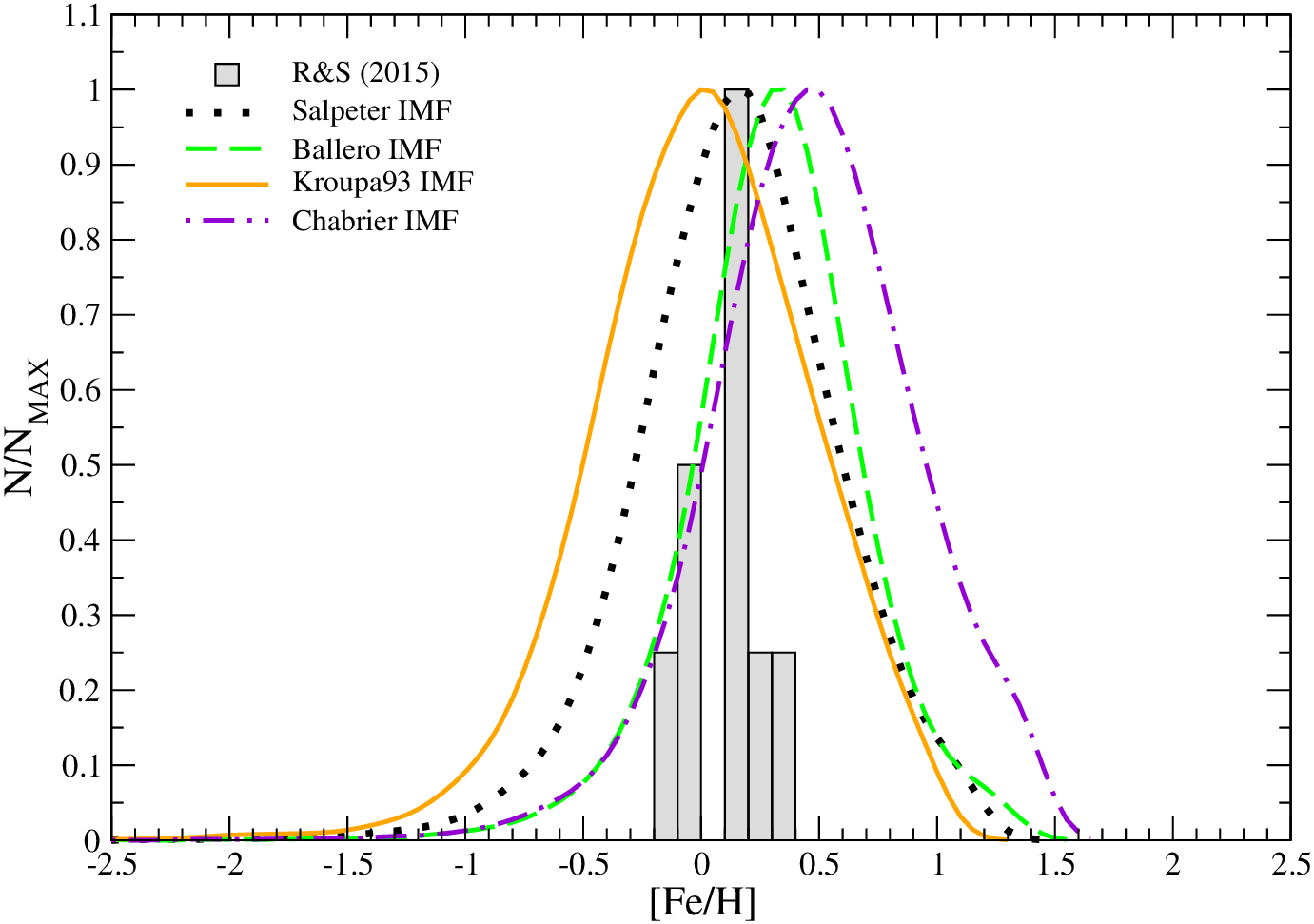} 
    \caption{Predicted MDF for  different models with star formation efficiency $\nu=25\, Gyr^{-1}$, infall timescale $T_{inf}=0.7\, Gyr$
    but with different IMFs; the lines have the same colour code of Figure \ref{MgSiCa_IMF}. 
    The data are from Ryde \& Schultheis (2015).  The predicted MDFs have been convolved with a Gaussian with an error of 0.1 dex.
    }
\label{MDF_IMF}
\end{figure*}

 The infall timescales also affect the MDF and in Figure \ref{MDF_Tinf} we show the predicted MDF for the Chabrier (2003) IMF and different infall timescales;
we note that for this IMF the infall timescale of  0.1 $Gyr$ produces results more in agreement with observations, while 
longer timescales tend to predict a peak at too high metallicities. Finally, in Figure \ref{MDF_IMF} we show the 
effect of varying the IMF on the predicted MDF, when the infall timescale is fixed at 0.7 $Gyr$. 
In this case, we note that the Salpeter IMF produces the best agreement and  predicts the peak of the MDF at 
the right position. However, given the small number of observed stars, we are not able to disentangle the degeneracy between IMF and infall timescale. 
What seems more clear is that an IMF like Kroupa et al. (1993), good for the solar vicinity, predicts the worst 
agreement with data when both the abundance ratios and the MDF in the GC are taken into account.
Previous models for the entire bulge (Brocato \& Matteucci, 1990; Ballero et al. 2007; Cescutti \& Matteucci, 2011; Kobayashi et al. 2011) 
have supported an IMF which contains more massive stars than the one for the solar vicinity.

In summary, the comparison between data and model results suggests that the abundances in the GC are quite compatible with a fast 
formation of this region similarly to what happened to the entire bulge. The fact that the [$\alpha$/Fe] ratios are low at high 
[Fe/H] is not necessarily a sign of prolonged star formation. Such a sign instead would be revealed by low  [$\alpha$/Fe] ratios 
at low [Fe/H], as happens in dwarf irregular galaxies and dwarf spheroidals where the SFR was quite mild.
A regime of low star formation, in fact, produces a situation where the gas is still poor in Fe when Type Ia SNe start producing 
the bulk of this element. At this point, the  [$\alpha$/Fe] ratios start to decrease but the [Fe/H] is still low. The fact that 
the star formation was quite strong in the early phases of formation of the central region is also evident in the MDF which is 
peaked at high metallicity. However, we should keep in mind that the number of stars observed in the GC is not yet statistically 
significant and therefore firm conclusions cannot be drawn. 
We recall that in this section we have ignored the fact that recent star formation has been observed in the GC region, so in the 
next section we will explore the case in which a second burst has recently occurred and whether this recent SFR can affect our predictions.

\subsection{Star Formation in the Galactic Center}
In our standard model for the GC, the bulk of the stars formed 10 $Gyr$ ago and then the SFR decreased steadily until the present 
time to a value slightly lower than $10^{-4} M_{\odot} yr^{-1}$.
Therefore, to reproduce the observations indicating active star formation at the present time, we have simulated a star formation 
history made up of a first strong burst, where most of stars formed (our fiducial model), plus a more recent starburst started 500 Myr 
ago. We have tested different cases and verified the effects of this second starburst on the [$\alpha$/Fe] ratios and the MDF.
\begin{figure*}
	  \centering   
    \includegraphics[scale=0.4]{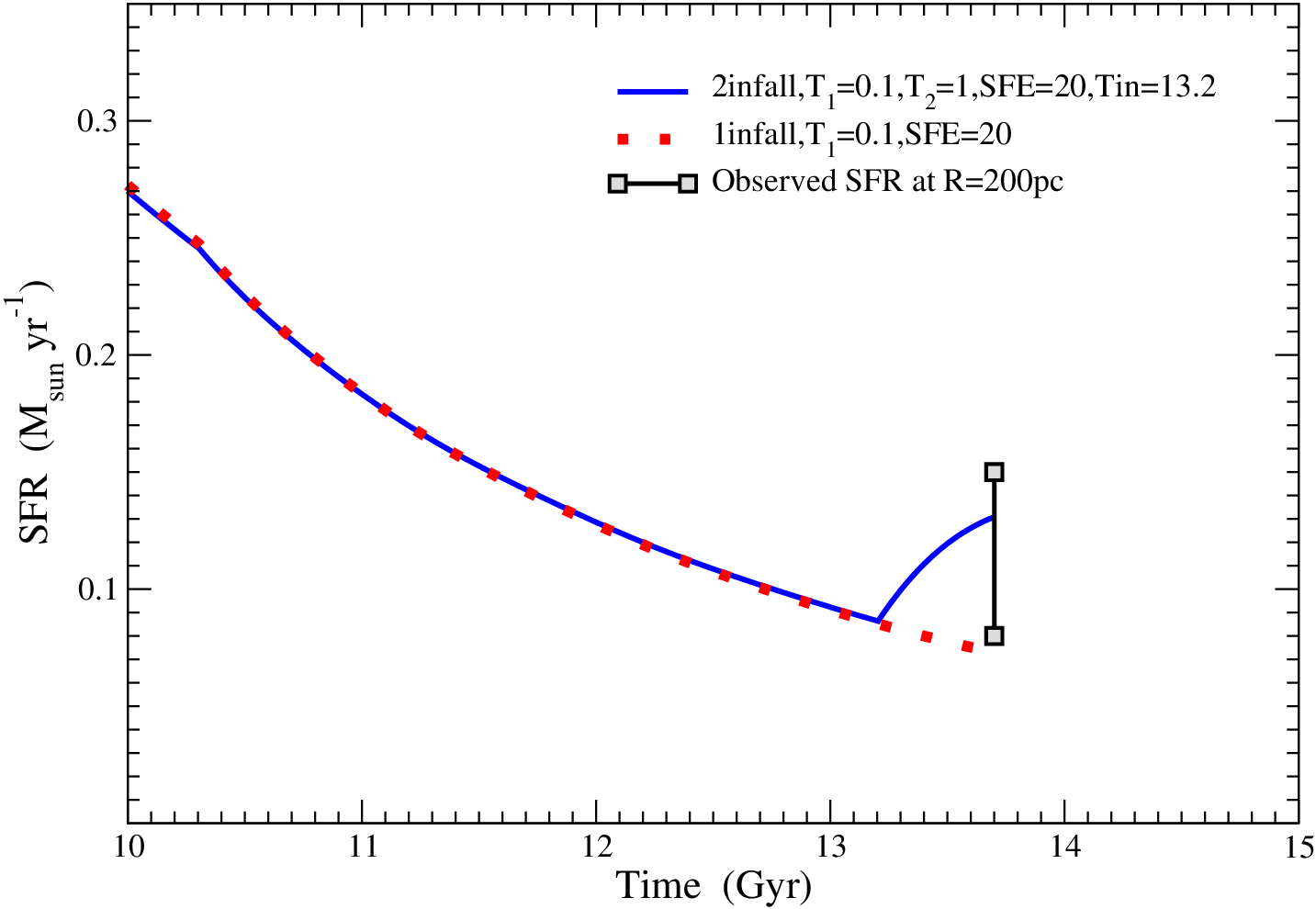} 
    \caption{The star formation history of the GC in the last few billion years. 
    The predicted star formation rate in $M_{\odot}yr^{-1}$ is given as a function of the galactic time in the case with a 
    second starburst starting 500 Myr ago (model ii continuous line). 
    The dashed red line indicates the model without any second burst of star formation. The adopted IMF in the models is that of Chabrier. 
    The observed minimum and maximum SFR in the GC are marked (see text).}
\label{SFR_obs}
\end{figure*}

\begin{figure*}
	  \centering   
    \includegraphics[scale=0.6]{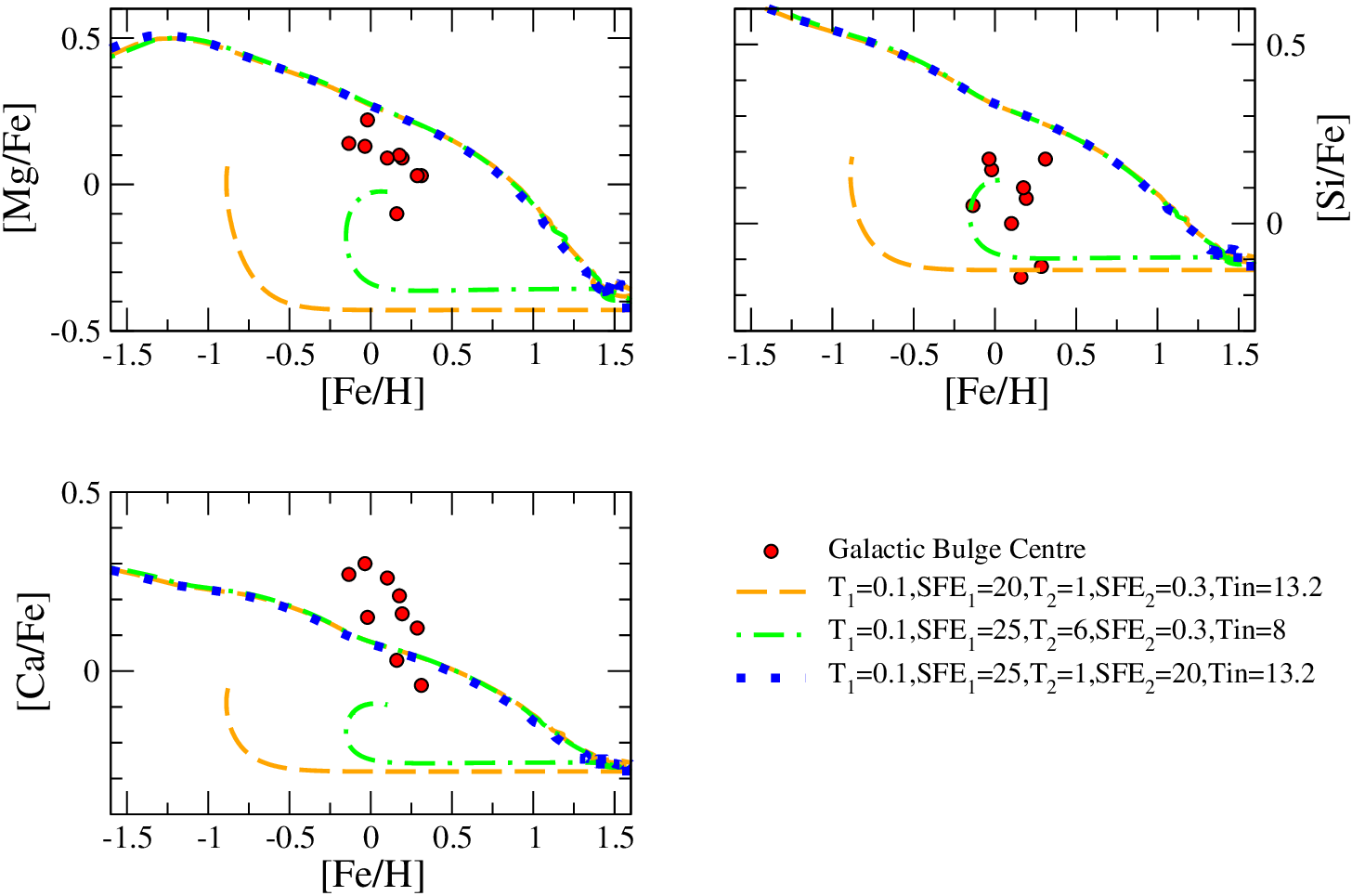} 
    \caption{Predicted and observed [$\alpha$/Fe] ratios for the following 
cases: the second burst of star formation triggered by a modest gas infall (model ii) indicated by the short-dashed blue line,  
strong infall of gas comparable to the initial one indicated by the long-dashed yellow line. Finally, the dashed-dotted 
green line indicates a model with the second burst triggered by a strong infall but starting at 8 instead of 13.2 $Gyr$. 
In the Figure are shown the parameters adopted in the the three models in order to obtain acceptable values for the SFR at the present time.
    }
\label{alpha_2infall}
\end{figure*}

In order to simulate the second burst, we simply assumed a second gas infall episode which increased suddenly the available mass of gas:
\begin{equation}
 \dot{\sigma}_{gas, i}(t)=A(r) X_{i, inf}e^{-t/T_{inf}}+B(r) X_{i, inf}e^{-(t-t_{in})/T_{inf_{2}}}
\end{equation}
where $t_{in}$ represents the cosmic time at which the second burst starts and 
$T_{inf_{2}}$ is the infall timescale of the second burst (see Table 1). The origin of this gas can be either extragalactic 
or coming from the inner regions of the Galactic disk thanks to the bar in the bulge.

\begin{table*}[htbp]
\begin{center}
\begin{tabular}{|c|c|c|c|c|c|}
\hline
Model  & $T_{inf} \,$($Gyr$) & $SFE_1 \,(Gyr^{-1})$ & $T_{inf_{2}} \,$($Gyr$)& $SFE_2 \, (Gyr^{-1})$ & $t_{in}$ ($Gyr$)  \\
\hline
i) &   0.1      &          20          &       1      &          0.3           &    13.2      \\
\hline
 ii) &      0.1      &          25          &       1      &          20           &    13.2      \\
\hline
iii) &      0.1      &          25          &       --      &          50           &     --\\
\hline
\end{tabular}
\end{center}
\caption{Parameters used for the two infall models:$T_{inf}$ and $T_{inf_{2}}$ are the infall timescale of the two SF episodes expressed in $Gyr$, 
$SFE_1$ and $SFE_2$ are the star formation efficiencies in unit of $Gyr^{-1}$ and finally $t_{in}$ is the time of the second infall onset.
For model iii) there is no second infall but just a sudden increase of the star formation efficiency passing from 25 to 50 $Gyr^{-1}$ at 13.2 $Gyr$.}
\label{2infall_parameters}
\end{table*}

We tested several cases but we show only the most realistic ones. In particular, we assumed: i) a burst triggered by a strong 
infall of gas, comparable to the initial infall rate, ii) a burst fueled by a more modest infall of gas relative to the 
previous one and iii) a starburst due only to a sudden increase of the star formation efficiency from 25 to 200 $Gyr^{-1}$ 
without an extra-infall of gas. This case could simulate the effect of a nuclear bar, which simply compresses the existing 
gas.  In the infall cases we tested both primordial and  metal enriched infall. We found that the best model is model 
ii) with a modest infall of new gas, which reproduces the estimated total mass of stars formed in the last 500 Myr ($\sim 10^{7} M_{\odot}$) 
as well as the observed present time SFR ($0.04-0.15M_{\odot} yr^{-1}$). 
In fact, in case i) the strong infall produces changes in the [$\alpha$/Fe] ratios which are not seen in the data: in particular, 
the infall dilutes the abundances of heavy elements thus producing an inversion in the  [$\alpha$/Fe] vs. [Fe/H] plot showing
very low [$\alpha$/Fe] ratios at low [Fe/H], not observed in the data. On the other hand, simply increasing the star formation 
efficiency (model iii)) does not allow us to reproduce the observed present time SFR nor the observed gas in the GC. 
This is because the gas was already low 500 Myr ago when the starburst was supposed to start. Therefore, we suggest that this 
burst in the GC must have been triggered by accretion of new gas. 
The characteristics of the  models i), ii) and iii) are reported in Table 1, where we show the timescale for the main infall 
episode ($T_{inf}$), the 
star formation efficiency for this episode ($SFE_1$), the timescale for the infall giving rise to the second burst ($T_{inf_{2}}$) 
and the star formation 
efficiency in this second burst ($SFE_2$). Finally, in column 6 the values of $t_{in}$  (the time at which the second infall starts) 
is reported. The best model appears to be model ii): it predicts a total mass of stars formed of $\sim 3.2 \cdot 10^{7}M_{\odot}$ and 
a present time SFR of $\sim 0.125 M_{\odot}yr^{-1}$, in good agreement with the observed values. 
In Figure \ref{SFR_obs} we show the predicted SFR history for model ii) and we report also the observations. 

In Figure \ref{alpha_2infall} we show the [$\alpha$/Fe] ratios  resulting from model i) and ii) which represent the case with 
a burst triggered by  a strong gas infall and a modest gas infall, respectively. 
In the same Figure is also shown an intermediate case showing that the inversion in the abundances, due to the dilution by infall,
is independent of the onset time of the second strong burst.
As one can see, a starburst triggered by a modest episode of primordial gas infall (at an average rate of $\sim 1.0 M_{\odot}yr^{-1}$), 
occurring in the last hundreds million years and reproducing the present time SFR, produces negligible effects on the 
predicted  [$\alpha$/Fe] ratios.
On the other hand, as already mentioned, a second starburst triggered by a strong infall of gas (with an initial rate of 
$\sim 200 M_{\odot}yr^{-1}$) would produce a noticeable but not observed effect in the data, even if the number of stars formed is the same. 
This case is clearly unrealistic since there is no physical justification for a strong late infall of primordial gas, but we show the 
resulting  [$\alpha$/Fe] ratios for the sake of comparison with case ii).
 Moreover, it is clear from Figure  \ref{alpha_2infall} that none of the models with a late strong infall fits the data.
We note that in all these cases we have chosen the model parameters to obtain 
acceptable values of the present time SFR as well as the total mass of stars formed. To do that we varied the star formation efficiencies 
and the infall timescale.
The reason for the inversion in the  [$\alpha$/Fe] ratios is therefore only due to the gas infall and to the way in which chemical 
elements are produced and on which timescales.
In particular, when a starburst is overimposed on a very low star formation regime, triggered by a huge episode of gas accretion, 
the abundances present at the moment in the ISM tend to be diluted by the 
infall, especially if the infalling gas is primordial,  but at the same time they increase  because of the new star formation. 
Therefore, according to the predominance of one or the other process, the abundances tend to increase or decrease. In our cases the abundances of Fe  
and the $\alpha$-elements initially tend to decrease because of the infall, thus producing a reverse trend at constant [$\alpha$/Fe], then 
as a result of the new production of $\alpha$-elements by Type II SNe, 
the [$\alpha$/Fe] 
ratios increase strongly. This is also due to the fact that Fe is not increasing immediately at the beginning of the second burst for the simple reason that 
Fe is mainly produced on long timescales by Type Ia SNe. No stars seem to be observed with the [$\alpha$/Fe] and [Fe/H] values predicted by the models with 
a late second starburst triggered by a huge gas infall.
Only a model with a second very early starburst would not affect the [$\alpha$/Fe] ratios, as expected. This fact suggests that a strong
burst like the one we have hypothesized in model i) cannot have occurred, otherwise it would have left a clear signature in the abundance 
pattern, and we have verified that this is true also for enriched infall. 
On the other hand, a minor burst triggered 
by either primordial or pre-enriched gas ( enriched in metals) would 
produce negligible effects on the stellar abundances, as we have shown in 
Figure 8. 
Concerning the MDF we tested model ii) and no differences were produced by the 
second infall.

 \section{Discussion and Conclusions}

In this paper we have studied the chemical evolution of the central region of the Galactic bulge (200 pc radius) 
and compared our results with the recent 
data from Ryde \& Schultheis (2015) concerning $\alpha$-elements and Fe and the present time star formation rate 
(Morris \& Serabyn 1996, Yusef-Zadeh et al. 2009, Molinari et al. 2011 and Immer et al. 2012).
We have tested different IMFs, efficiencies of star formation and infall timescales. We also assumed a recent and 
ongoing burst of star formation where $\sim 3.2\cdot 10^{7}M_{\odot}$ of new stars formed.

Our main conclusions can be summarized as follows:
 \begin{itemize}
 \item A comparison between data and models suggests that the [$\alpha$(Mg,Si,Ca)/Fe]  ratios in the GC can be 
 reproduced by a model assuming a strong star formation at the beginning such as that assumed for the whole bulge 
 by previous models (Matteucci \& Brocato, 1990; Ballero et al. 2007; Cescutti \& Matteucci 2011; Grieco et al. 2012). Under this assumption, most of the 
 stars in the GC have formed at the beginning (more than 10 $Gyr$ ago) and the SFR has decreased steadily until the 
 present time.

 \item For [Mg/Fe] and [Si/Fe] we can conclude that a classical Salpeter IMF can reproduce the data, whereas for Ca the 
 data always lie above the predictions and this is true also for the other assumed IMFs. We conclude that for Ca there could be
a problem with stellar yields. In fact, stellar yields of some elements are still uncertain: here, we have adopted 
 the set of stellar yields, for stars of all masses, considered the best to reproduce the abundance patterns in the 
 solar vicinity (Romano et al. 2010, where also a discussion on the uncertainties on stellar yields can be found).
 
 \item We have varied the efficiency of star formation from 20 $Gyr^{-1}$ to 200 $Gyr^{-1}$ and found practically no 
 difference in the results concerning the [$\alpha$/Fe] vs.[Fe/H] relations, whereas the MDF allows us to exclude the efficiency 
 of 200 $Gyr^{-1}$. Therefore, we conclude that the efficiency of star formation in the GC is similar to that inferred 
 for the whole bulge (GMPC12), namely 20-25 $Gyr^{-1}$.
 
 \item We have tested different IMFs: Salpeter, Chabrier, Ballero and Kroupa93 and found that  Salpeter,  Ballero and Chabrier IMFs 
 are probably the best to describe the GC, 
 in agreement with previous results for the whole bulge suggesting an IMF with more massive stars than in
 the solar vicinity (Matteucci\& Brocato, 1990; Ballero et al. 2007; Cescutti \& Matteucci, 2011; Grieco et al. 2012); 
 in fact, they produce the best fits of the [$\alpha$/Fe] vs. [Fe/H] relations as well as the MDF, at the same time. 
 However, the Salpeter IMF requires an infall
 timescale longer (0.7 $Gyr$) than the other two IMFs (0.1 $Gyr$) to fit the data. Given the uncertainties still existing 
 in the data we are not able at the moment to solve this degeneracy and therefore to draw firm conclusions on this 
point, but we can say that the infall timescale for the formation of the bulk of GC stars was short.
 
\item We tested also the effect of changing the infall timescale and varied this timescale from 0.01 $Gyr$ (practically a 
closed-box model) to 1.25 $Gyr$. Again, with the exception of Ca, for which all the model predictions lie below the data, 
the Mg and Si data are compatible with timescales in the range 0.1-1.25 $Gyr$,  whereas the MDF is compatible with a range 0.1-07 $Gyr$.

\item Finally, we tested the effects of a second very recent burst occurring in the last five hundreds million years and 
having a SFR and a total amount of formed stars comparable to those observed. We have found that such a burst, if 
triggered by a new modest infall episode of gas of primordial composition, would not produce any noticeable 
effect in the  [$\alpha$/Fe] relations and the MDF.
On the other hand, we tested also a case of a starburst triggered by a huge gas infall, still producing the same present 
time SFR and total mass of new stars, and found that in this case the [$\alpha$/Fe] are strongly affected by the infall. 
In particular, we predict very low values of [$\alpha$/Fe] ratios at low [Fe/H], which are not observed.

\item Therefore, we conclude that the stars in the GC formed very quickly with a star formation efficiency of $\sim 25 \,Gyr^{-1}$, and
that the IMF should contain a larger number of massive stars than the typical IMF for the solar vicinity, such as 
the Kroupa et al. (1993). In particular, the recent IMF of Chabrier (2003) seems to be preferred. The present time 
observed SFR in the GC can be explained by a second burst of star formation with the same efficiency as the previous 
one and the same IMF, triggered by a modest infall episode. This second burst would not produce evident effects on the 
abundance patterns nor on the MDF, and is compatible with gas accretion induced by the main bar in the bulge.
 
 \end{itemize}

\section*{Acknowledgments}
 We acknowledge financial support
from PRIN MIUR 2010-2011, project \textquotedblleft The Chemical and dynamical
Evolution of the Milky Way and Local Group Galaxies\textquotedblright, prot.
2010LY5N2T.
NR acknowledges support from the Swedish Research Council,
VR (project number 621-2014-5640), 
Funds from Kungl. Fysiografiska S\"allskapet i Lund. 
(Stiftelsen Walter Gyllenbergs fond and M\"arta och Erik Holmbergs donation). 
 We thank I.J. Danziger for reading the manuscript. Finally, we thank an anonymous referee for his/her 
 important suggestions which improved the paper.

\end{document}